\documentclass[fleqn,usenatbib]{mnras}

\usepackage{newtxtext,newtxmath}

\usepackage[T1]{fontenc}

\DeclareRobustCommand{\VAN}[3]{#2}
\let\VANthebibliography\thebibliography
\def\thebibliography{\DeclareRobustCommand{\VAN}[3]{##3}\VANthebibliography}

\usepackage[dvipdfmx]{graphicx}	
\usepackage{amsmath}	
\usepackage{bm}  
\usepackage{color}
\usepackage{ulem}

\definecolor{ao_en}{rgb}{0.0, 0.5, 0.0}

\defcitealias{2021MNRAS.506.5512F}{Paper I}

\title[GC formation with top-heavy IMF]{
The formation of globular clusters with top-heavy initial mass functions
}

\author[H.Fukushima and H.Yajima]{
Hajime Fukushima$^{1}$\thanks{E-mail:fukushima@ccs.tsukuba.ac.jp},
Hidenobu Yajima$^{1}$
\\
$^{1}$Center for Computational Sciences, University of Tsukuba, Ten-nodai, 1-1-1 Tsukuba, Ibaraki 305-8577, Japan\\
}
\date{Accepted XXX. Received YYY; in original form ZZZ}

\pubyear{2020}

\begin{document}
\label{firstpage}
\pagerange{\pageref{firstpage}--\pageref{lastpage}}
\maketitle

\begin{abstract}
We study the formation of globular clusters in massive compact clouds with the low-metallicity of $Z=10^{-3}~Z_{\odot}$ by performing three-dimensional radiative-hydrodynamics simulations. 
Considering the uncertainty of the initial mass function (IMF) of stars formed in low-metallicity and high-density clouds, we investigate the impacts of the IMF on the cloud condition for the GC formation with the range of the power-law index of IMF as $\gamma = 1-2.35$. We find that the threshold surface density ($\Sigma_{\rm thr}$) for the GC formation increases from $800~M_{\odot} \; {\rm pc^{-2}}$ at $\gamma = 2.35$ to $1600~M_{\odot}\; {\rm pc^{-2}}$ at $\gamma = 1.5$ in the cases of clouds with $M_{\rm cl} = 10^6~M_{\odot}$ because the emissivity of ionizing photons per stellar mass increases as $\gamma$ decreases. For $\gamma 
 < 1.5$, $\Sigma_{\rm thr}$ saturates with $\sim 2000~M_{\odot}\; {\rm pc^{-2}}$ that is quite rare and observed only in local starburst galaxies due to e.g., merger processes.
Thus, we suggest that formation sites of low-metallicity GCs could be limited only in the very high-surface density regions. We also find that $\Sigma_{\rm thr}$ can be modelled by a power-law function with the cloud mass ($M_{\rm cl}$) and the emissivity of ionizing photons ($s_*$) as $\propto M_{\rm cl}^{-1/5} s_{*}^{2/5}$.
Based on the relation between the power-law slope of IMF and $\Sigma_{\rm thr}$, future observations with e.g., the James Webb Space Telescope can allow us to constrain the IMF of GCs.
\end{abstract}

\begin{keywords}
stars:formation - stars:massive - stars:Population II - H{\sc ii} regions - galaxies: star clusters: general - galaxies: star formation.
\end{keywords}


\section{Introduction}\label{introduction} 
Globular clusters (GCs) are survivors of ancient star clusters born in the early Universe from the estimated age $\gtrsim 10~{\rm Gyr}$ \citep[e.g.,][]{2010ARA&A..48..431P}.
GCs can be keys for understanding cosmic star formation rate and galaxy evolution in the early Universe.
However, the formation mechanism of GCs has not been understood yet. 
Previous studies with cosmological simulations showed that massive and dense star clusters are formed in the early galaxies \citep{2016ApJ...831..204R, 2018MNRAS.475.4252A, 2021MNRAS.500.4578M}.
In observations, \citet{2017MNRAS.467.4304V} discovered the young stellar system heavier than $10^5~M_{\odot}$ at $z\sim 6$ \citep[see also][]{2019MNRAS.483.3618V}.
Very recently, James Webb Space Telescope (JWST) has allowed us to probe massive and compact star clusters more precisely. 
\citet{2022arXiv220800520V} discovered the star cluster whose stellar mass is comparable to GCs ($\sim 10^6~M_{\odot}$) at $z\sim 4$.

In the GC formation, efficient conversion from the gas to stars is a necessary condition. After evaporation of a star-forming cloud, the stellar density decreases significantly if the stellar system is gravitationally unbound.
$N$-body simulations showed that 
more than 10-30 percent of the gas should be converted into stars to make the star cluster gravitationally bound state 
\citep[e.g.,][]{1984ApJ...285..141L, 2001MNRAS.321..699K, 2007MNRAS.380.1589B, 2017A&A...605A.119S}.
Recent hydrodynamics simulations show that the bound fractions of star clusters are tightly related to the star formation efficiencies (SFEs)  \citep{2019MNRAS.487..364L, 2021MNRAS.506.3239G, 2022MNRAS.511.3346F}.
In the Milky Way, the SFEs are typically less than 10 percent \citep[e.g.,][]{1986ApJ...301..398M, 2020MNRAS.493.2872C}, and so most star clusters are dispersed after their birth \citep{2003ARA&A..41...57L}.

In star cluster formation, massive stars disrupt their host clouds with stellar feedback, such as radiative feedback, stellar wind, and supernovae (SNe) \citep[e.g.,][]{2019ARA&A..57..227K, 2022arXiv220309570C}.
Extreme ultraviolet (EUV; $13.6~{\rm eV} \lesssim h \nu \lesssim 1~{\rm keV}$ ) photons ionize the hydrogen atoms and heat gas to $10^4~{\rm K}$ of which the high thermal pressure can disrupt clouds
\citep[e.g.,][]{1997ApJ...476..166W, 2002ApJ...566..302M, 2009ApJ...703.1352K, 2010ApJ...710L.142F, 2016ApJ...819..137K, 2020MNRAS.497.5061I}.
This photoionization feedback can regulate the SFEs to be less than $\sim 0.1$ in the environments of the Milky Way
\citep[e.g.,][]{2010ApJ...715.1302V, 2012MNRAS.427.2852D, 2013MNRAS.430..234D, 2017MNRAS.470.3346H,2017MNRAS.471.4844G, 2017MNRAS.472.4155G, 2018ApJ...859...68K, 2019MNRAS.489.1880H, 2020MNRAS.497.4718D, 2018MNRAS.475.3511G, 2020MNRAS.497.3830F, 2021MNRAS.506.3239G, 2020MNRAS.499..668G,2020MNRAS.495.1672B, 2021MNRAS.501.4136A, 2021PASJ..tmp...65F, 2022MNRAS.509..954D, 2022ApJS..259...21K, 2022MNRAS.512..216G}.
Stellar wind also injects kinetic energy into the ambient matters \citep{1977ApJ...218..377W}.
However, \citet{2021ApJ...914...89L} showed that the mixing layers around the bubble efficiently lost energy, resulting in the weak impacts on the cloud disruption \citep[see also][]{2021ApJ...914...90L, 2021ApJ...922L...3L}.
Supernova feedback is powerful enough to evacuate the gas from a cloud \citep[e.g.,][]{2016MNRAS.463.3129G}.
Yet, recent observations indicate that clouds are disrupted before the end of the lifetime of OB stars \citep{2019Natur.569..519K}.
Thus, photoionization is likely to play a primary role in the star cluster formation \citep{2022arXiv220510413G}.
Note that the radiation pressure on dust grains also regulates the star formation \citep{2010ApJ...709..191M, 2015ApJ...809..187S, 2016ApJ...829..130R}.
However, it should be secondary because of the low-metallicities of clouds hosting GCs.

In understanding the relationship between the photoionization feedback and the star cluster formation, the initial mass function (IMF) is an essential factor. The production rate of ionizing photons sensitively depends on the shape of the IMF.
It is well known that the IMF in the nearby star-forming regions is universal \citep[e.g.][]{1955ApJ...121..161S, 2002Sci...295...82K, 2003PASP..115..763C}.  
Observations of massive star clusters indicated that the IMF is hop-heavy in these highly dense clusters \citep{2013ApJ...764..155L, 2016MNRAS.458.3027M, 2018Sci...359...69S, 2019ApJ...870...44H}.
The observed slopes of the IMFs are $\gamma = 1.7-1.9$, and there is an excess of massive stars.
\citet{2022arXiv220303276P} showed that the core mass function is shallower than the Salpeter IMF in the massive star-forming regions ($\gamma = 1.95$, W43-MM2\&MM3). 
Besides, indirect evidence of the top-heavy IMF in GCs is also found \citep{2012MNRAS.422.2246M}.
They found that the IMF of GCs with low-metallicities or high-stellar densities tends to be top-heavy. 
Using their results, \citet{2016ApJ...826...89Z} showed that the top-heavy IMF could nicely explain the relation between the mass-to-light ratios and metallicity of GCs in M31 \citep[see also,][]{2017ApJ...839...60H}.
On the other hand, \citet{2023arXiv230301636B} indicated that there is no evidence of top-heavy IMFs in GCs of the Galaxy and the Large/Small Magellanic Clouds in the metallicity range of $Z>10^{-2}~Z_{\odot}$.
Therefore, the IMF of low-metallicity GC is still under debate.
In particular, the case with a metallicity lower than $\sim 10^{-2}~Z_{\odot}$  is poorly understood.

Recently, \citet{2021MNRAS.508.4175C} performed hydrodynamics simulations, including the cooling effects of metal lines and dust grains.
They showed that the highly filamentary structures are induced, and the Chabrier-like IMF \citep{2003PASP..115..763C} is realized at $Z \gtrsim 10^{-1}~Z_{\odot}$.
In the low-metallicity environments with $Z\lesssim 10^{-2}~Z_{\odot}$, the formation of the filamentary structures is suppressed, resulting in the top-heavy IMF. 
Also, \citet{2022MNRAS.512.2573T} indicated that the IMF should be top-heavy with $\gamma < 1 $ around the active galactic nuclei to reproduce the observed Fe {\sc ii} / Mg {\sc ii} line-flux ratio.
Thus, the IMF is likely top-heavy in the low metallicity and the high gas density environments.

In this paper, we study the conditions of clouds for GC formation with top-heavy IMFs. We consider clouds with
 masses of $M_{\rm cl} = 10^6$ and $10^7~M_{\odot}$ and the cloud surface densities of $\Sigma_{\rm cl} \sim 400-2000~M_{\odot}{\rm pc^{-2}}$.
We utilize the 3D radiative-hydrodynamics (RHD) simulation code, \textsc{SFUMATO-M1} which is the modified code of a self-gravitational magnetohydrodynamics code with an Eulerian adaptive mesh refinement (AMR), \textsc{sfumato} \citep{2007PASJ...59..905M, 2015ApJ...801...77M}.
We adopt the radiation transfer scheme based on the momentum method with the M1-closure and the stochastic stellar population developed in \citet[][hereafter \citetalias{2021MNRAS.506.5512F}]{2021MNRAS.506.5512F} and \citet{2022MNRAS.511.3346F}.

We organize the rest of this paper as the following.
In Section \ref{Sec_simulation_method}, we describe the numerical method and the initial conditions of the simulations.
Then, we show the results of the simulations in Section \ref{Sec_results}.
In Section \ref{Sec:discussion}, we discuss the implications of our simulations on the low-metallicity GC formation.
Section \ref{Sec:summary} is for the summary.

\section{Simulation Method}\label{Sec_simulation_method}

We perform the RHD simulations with {\sc SFUMATO-M1} \citepalias{2021MNRAS.506.5512F}, the modified version of self-gravitational magnetohydrodynamics code {\sc sfumato} \citep{2007PASJ...59..905M} which utilizes the adaptive mesh refinement technique.
Our simulations include on-the-fly radiative transfer calculations with the M1-closure technique.
As in \citet{2013MNRAS.436.2188R}, we adopt the approximation with the reduced speed of light for the radiation transfer with $\tilde{c} = 3 \times 10^{-4} c$ where $c$ is the speed of light.
We set the four frequency bins for (1) extreme ultraviolet ($13.6~{\rm eV}< h\nu$), (2) Lyman-Werner ($11.2~{\rm eV} < h \nu < 13.6~{\rm eV}$), (3) far-ultraviolet ($6~{\rm eV} < h \nu < 13.6~{\rm eV}$), and (4) infrared photons.
The chemistry solver is developed in \citet{2020ApJ...892L..14S} which studied the primordial star formation.
We added the chemical network of CO and the oxygen ion in the H{\sc ii} regions \citep[O{\sc ii} and O{\sc iii},][]{2020MNRAS.497..829F}.
We also adopt the sink particle technique for modelling a star cluster as in \citet{2015ApJ...801...77M}.
Further detail of the simulation methods is described in \citetalias{2021MNRAS.506.5512F} and \citet{2022MNRAS.511.3346F}. 
In this study, we fixed the metallicity at $Z=10^{-3}~Z_{\odot}$.

\subsection{Model of stellar population}\label{sec:stellar_population}

\begin{figure}
    \begin{center}
    	\includegraphics[width=\columnwidth]{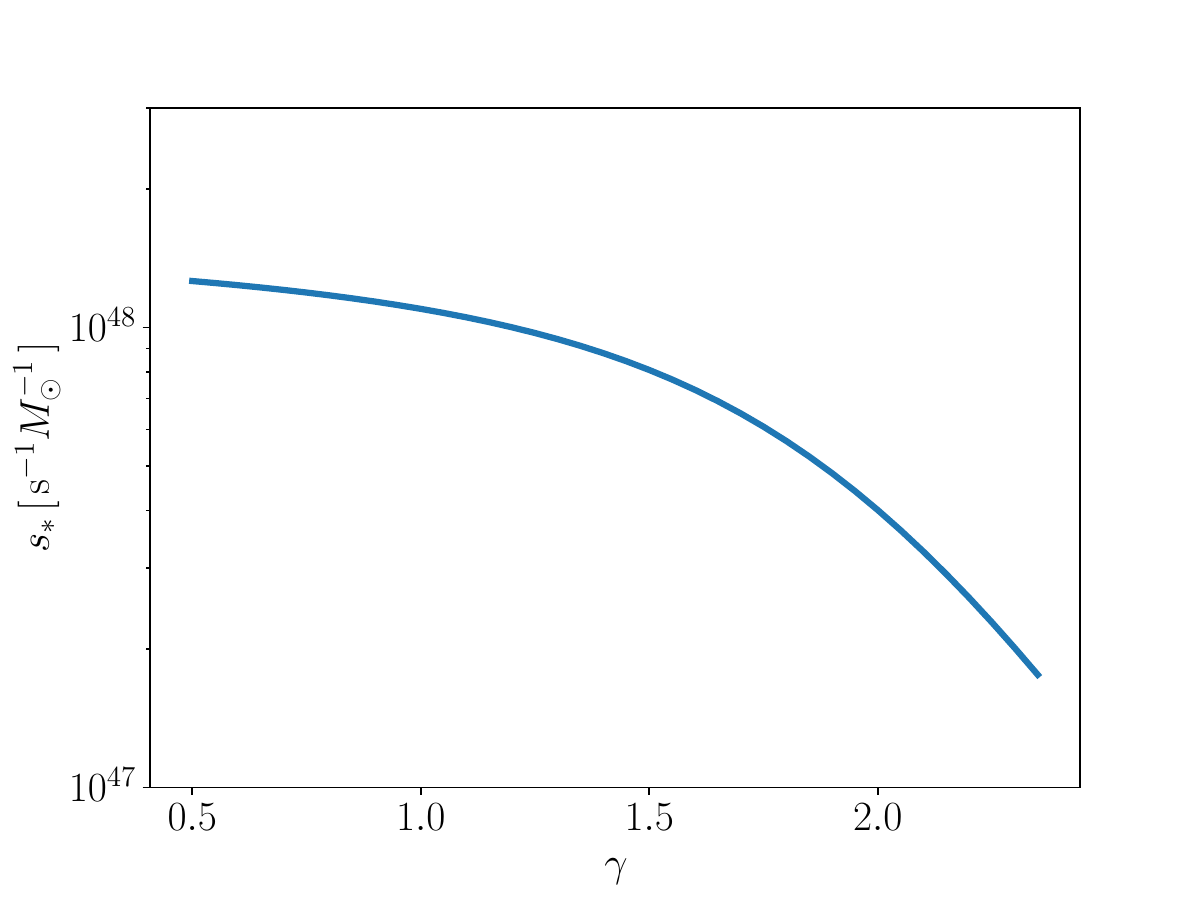}
    \end{center}
    \caption{
    The emissivity of ionizing photons as a function of the slope of the stellar IMF.
}   
    \label{fig:sstar_ion_gam}
\end{figure}

We adopt the stochastic stellar population model developed in \citet{2022MNRAS.511.3346F}.
This model is similar to {\sc SLUG} code developed by \citet{2012ApJ...745..145D} and \citet{2015MNRAS.452.1447K}.
We divide the masses into 150 bins between $0.1$ and $300~M_{\odot}$.
In the Milky way, it is known that there is an upper limit in stellar mass around $150~M_{\odot}$ \citep{2004MNRAS.348..187W, 2005Natur.434..192F}.
However, a massive star whose initial mass is more than $\sim 200~M_{\odot}$ was discovered in the R136 star cluster in the Large Magellanic Cloud \citep{2010MNRAS.408..731C, 2022arXiv220713078K}.
The upper stellar mass in the GCs is still unknown. Thus, we adopt $300~M_{\odot}$ as the upper stellar mass limit.
The newborn stars are stochastically distributed into bins based on the probability from the IMF.
We adopt the Chabrier IMF \citep{2003PASP..115..763C} but  change the slope of the massive end.
In the standard IMF, the slope is $\gamma = 2.35$ as the Salpeter IMF \citep{1955ApJ...121..161S}, where the number of stars within a specific mass range is $dn \propto m^{-\gamma} dm$.
Figure \ref{fig:sstar_ion_gam} shows  the emissivity of ionizing photons per unit stellar mass averaged over the IMF as a function of $\gamma$.
The shape of the IMF in a low-metallicity massive star cluster is still unknown.
\citet{2021MNRAS.508.4175C} indicated that the slope of IMF is shallower than $\gamma = 2$.
Thus, we mainly consider the case with $\gamma = 1.5$ as the fiducial model of the top-heavy IMF.
Besides, we calculate the models with $\gamma = 1$ and $2$.
In the low-mass star clusters ($M_*<10^4~M_{\odot}$), the emissivity is not uniform due to the stochastic stellar population \citep{2016ApJ...819..137K} even if the total stellar mass is same. 
However, the total stellar masses in our simulations are larger than $10^{4}~M_{\odot}$.
Therefore, the radiative properties of GCs are almost the same as the IMF averaged values.

The stars are distributed in each sink particle at different times, and their radiative properties depend on their ages.
We consider the stellar evolution to estimate their radiative properties by adopting the {\sc PARSEC} tracks \citep{2012MNRAS.427..127B, 2014MNRAS.445.4287T, 2014MNRAS.444.2525C, 2015MNRAS.452.1068C, 2017ApJ...835...77M, 2019MNRAS.485.5666P, 2020MNRAS.498.3283P}.
The lower metallicity limit of {\sc PARSEC} tracks is $Z=10^{-2}~Z_{\odot}$, and thus we adopt the stellar evolution track at this value.
In addition, we use the SED models in \citet{1997A&AS..125..229L} and \citet{2019A&A...621A..85H} for OB-stars in SMC to calculate the emissivity of ionizing photons.
We note that the typical mass ranges of sink particles are $10^2-10^4~M_{\odot}$ ($10^3-10^5~M_{\odot}$), and its mean values are $\sim 10^3~M_{\odot}$ ($\sim 10^4~M_{\odot}$) with the cloud mass $10^6~M_{\odot}$ ($10^7~M_{\odot}$) in our simulations.

\begin{table}
    \caption{Models considered.}
    \label{Tab:models}
    \centering
    \begin{tabular}{|l|c|c|c|c|c|c|c|}
        \hline \hline
        model & $M_{\rm cl} $  & $\Sigma_{\rm cl}$ & $R_{\rm cl}$ & $\gamma$ & $t_{\rm ff}$ & $v_{\rm esc} $  \\
        & $[M_{\odot} ]$  & $[M_{\odot} {\rm pc^{-2}}]$ & $[{\rm pc}]$  & $ [{\rm Myr} ]$ & $[{\rm km/s} ]$ \\
        \hline
M6R125G15 & $10^6$ & $2000$ & $12.5$ &  $1.5$ & $0.73$ & $26$ \\
M6R13G1 & $10^6$ & $2000$ & $13$ & $1$ & $0.74$ & $26$ \\
M6R135G15 & $10^6$ & $1700$ & $13.5$ & $1.5$ & $0.82$ & $25$ \\
M6R14G15 & $10^6$ & $1600$ &$14$  & $1.5$ & $0.87$ & $25$ \\
M6R15G15 & $10^6$ & $1400$ &$15$  & $1.5$ & $0.96$ & $24$ \\
M6R14G1 & $10^6$ & $1700$ &$14$  & $1$ & $0.84$ & $25$ \\
M6R16G2 & $10^6$ & $1300$ & $16$ & $2$ & $1.0$ & $23$ \\
M6R18G2 & $10^6$ & $1000$ &$18$  & $2$ & $1.2$ & $22$ \\
M6R19G2 & $10^6$ & $900$ & $19$  & $2$ & $1.3$ & $21$ \\
M6R20G15 & $10^6$ & $800$ & $20$  & $1.5$ & $1.5$ & $21$ \\
M6R20G235 & $10^6$ & $800$ &$20$  & $2.35$ & $1.5$ & $21$ \\
M6R23G235 & $10^6$ & $600$ & $23$ & $2.35$ & $1.8$ & $19$ \\
M6R28G235 & $10^6$ & $410$ & $28$  & $2.35$ & $2.5$ & $18$ \\
M7R46G15 & $10^7$ & $1500$ & $46$  & $1.5$ & $1.6$ & $43$ \\
M7R50G15 & $10^7$ & $1300$ & $50$ & $1.5$ & $1.9$ & $41$ \\
M7R56G15 & $10^7$ & $1000$ &$56$  & $1.5$ & $2.2$ & $39$ \\
M7R63G15 & $10^7$ & $800$ & $63$ & $1.5$ & $2.6$ & $37$ \\
M7R63G235 & $10^7$ & $800$ & $63$ & $2.35$ & $2.6$ & $37$ \\
M7R73G235 & $10^7$ & $600$ & $73$  & $2.35$ & $3.3$ & $34$ \\
M7R89G235 & $10^7$ & $400$ & $89$ & $2.35$ & $4.4$ & $31$ \\
        \hline
    \end{tabular}
    \begin{minipage}{1 \hsize}
    Notes. Column 1: model names, Column 2: cloud masses, Column 3: cloud radii, Column 4: surface densities, Column 5: high-mass IMF slope, Column 6: free-fall times, Column 7: escape velocities.
    \end{minipage}
\end{table}

\subsection{Initial condition}\label{Sec:Initial_condition}
We perform the simulations of clouds with the masses $M_{\rm cl} = 10^6~M_{\odot}$ and $10^{7}~M_{\odot}$.
We change the slopes of the IMF as $\gamma = 1$, $1.5$, $2$, and $2.35$.
By considering the variety of initial states of the clouds, we investigate the cases with $\Sigma_{\rm cl} =400-2000~M_{\odot}{\rm pc^{-2}}$.
The models are summarized in Table \ref{Tab:models}.
Hereafter, we label models according to which the cloud mass ($M_{\rm cl}$), the radius ($R_{\rm cl}$), and the slope of the IMF ($\gamma$).
For example, "M6R20G235" represents the model with $M_{\rm cl}=10^6~M_{\odot}$, $R_{\rm cl} = 20~{\rm pc}$, and $\gamma = 2.35$.
In each simulation, the maximum refinement level is fixed at $l_{\rm max} = 4$.
The minimum cell size is $\Delta x = 0.059 ~{\rm pc} (R_{\rm cl}/20~{\rm pc})$ where $R_{\rm cl}$ is the cloud radius. 
Our simulations can resolve compact HII regions in high-density regions in a cloud sufficiently.
As in \citetalias{2021MNRAS.506.5512F}, we set the turbulent motions in the initial condition with the velocity power spectrum as $P(k) \propto k^{-4}$ where $k$ is the wavenumber.
The strength of the turbulent motions is characterized by the virial parameter defined as 
\begin{align}
    \alpha_{0} = \frac{2E_{\rm kin}}{ | E_{\rm grav}|} = \frac{5 \sigma_0^2 R_{\rm cl}}{3 G M_{\rm cl}}, \label{eq_alpha_vir}
\end{align}
where $\sigma_0$, $E_{\rm kin}$, and $E_{\rm grav}$ are the 3D velocity dispersion, kinetic and gravitational energy.
Here, we adopt $\alpha_0 = 1$.

\section{Results} \label{Sec_results}

We first represent the impact of the top-heavy IMF on the dense star cluster formation in Section \ref{Sec:star_cluster_formation}.
In Section \ref{Sec:dependence_on_cloud_compact}, we investigate the condition for the formation of dense star clusters.
In Table \ref{Tab:results}, we summarize the results of our simulations.

\begin{table}
    \caption{Simulation results.}
    \label{Tab:results}
    \centering
    \begin{tabular}{|l|c|c|c|c|c|}
        \hline \hline
        model & $\varepsilon_*$ & $M_{\rm bd}$ & $f_{\rm bd}$ & $r_{\rm h}$ & $\rho_*$ \\
              &                 & $[\, M_{\odot} \,]$ &       & $[\, \rm pc \,]$ & $[\, M_{\odot} {\rm pc^{-3}} \,]$ \\
        \hline
M6R125G15 & $0.40$ & $3.9 \times 10^5$ & $0.98$ & $0.22$ & $9.1 \times 10^6$ \\
M6R13G1 & $0.18$ & $1.3 \times 10^5$ & $0.73$ & $1.8$ & $5.2 \times 10^3$ \\
M6R135G15 & $0.31$ & $3.0 \times 10^5$ & $0.96$ & $0.48$ & $6.5 \times 10^5$ \\
M6R14G15 & $0.18$ & $1.3 \times 10^5$ & $0.74$ & $1.9$ & $4.7 \times 10^3$ \\
M6R15G15 & $0.15$ & $7.8 \times 10^4$ & $0.53$ & $3.4$ & $5.0 \times 10^2$ \\
M6R14G1 & $0.13$ & $5.0 \times 10^4$ & $0.38$ & $2.8$ & $5.7 \times 10^2$ \\
M6R16G2 & $0.39$ & $3.8 \times 10^5$ & $0.98$ & $0.31$ & $3.0 \times 10^6$ \\
M6R18G2 & $0.19$ & $1.6 \times 10^5$ & $0.84$ & $2.8$ & $1.7 \times 10^3$ \\
M6R19G2 & $0.16$ & $1.2 \times 10^5$ & $0.75$ & $3.8$ & $5.1 \times 10^2$ \\
M6R20G15 & $0.078$ & $8.2 \times 10^3$ & $0.11$ & $1.5$ & $6.3 \times 10^2$ \\
M6R20G235 & $0.49$ & $4.8 \times 10^5$ & $0.99$ & $0.39$ & $2.0 \times 10^6$ \\
M6R23G235 & $0.17$ & $1.2 \times 10^5$ & $0.73$ & $5.1$ & $2.2 \times 10^2$ \\
M6R28G235 & $0.11$ & $2.6 \times 10^4$ & $0.23$ & $6.8$ & $1.9 \times 10^1$ \\
M7R46G15 & $0.48$ & $4.5 \times 10^6$ & $0.99$ & $0.91$ & $1.4 \times 10^6$ \\
M7R50G15 & $0.32$ & $2.9 \times 10^6$ & $0.96$ & $1.3$ & $3.5 \times 10^5$ \\
M7R56G15 & $0.11$ & $5.9 \times 10^5$ & $0.54$ & $12.0$ & $7.8 \times 10^1$ \\
M7R63G15 & $0.076$ & $8.3 \times 10^4$ & $0.11$ & $2.2$ & $2.4 \times 10^3$ \\
M7R63G235 & $0.60$ & $5.8 \times 10^6$ & $1.0$ & $1.4$ & $4.6 \times 10^5$ \\
M7R73G235 & $0.45$ & $4.4 \times 10^6$ & $0.99$ & $1.9$ & $1.6 \times 10^5$ \\
M7R89G235 & $0.13$ & $5.6 \times 10^5$ & $0.43$ & $19.0$ & $1.9 \times 10^1$ \\
      \hline
    \end{tabular}
    \begin{minipage}{1 \hsize}
    Notes. Column 1: model names, Column 2: star formation efficiencies, Column 3: gravitationally bound masses, Column 4: gravitational bound fraction of star clusters, Column 4: half-mass radii, Column 5: stellar densities of star clusters.
    \end{minipage}
\end{table}

\subsection{Star cluster formation with top-heavy IMF}\label{Sec:star_cluster_formation}

\begin{figure*}
    \begin{center}
    	\includegraphics[width=185mm]{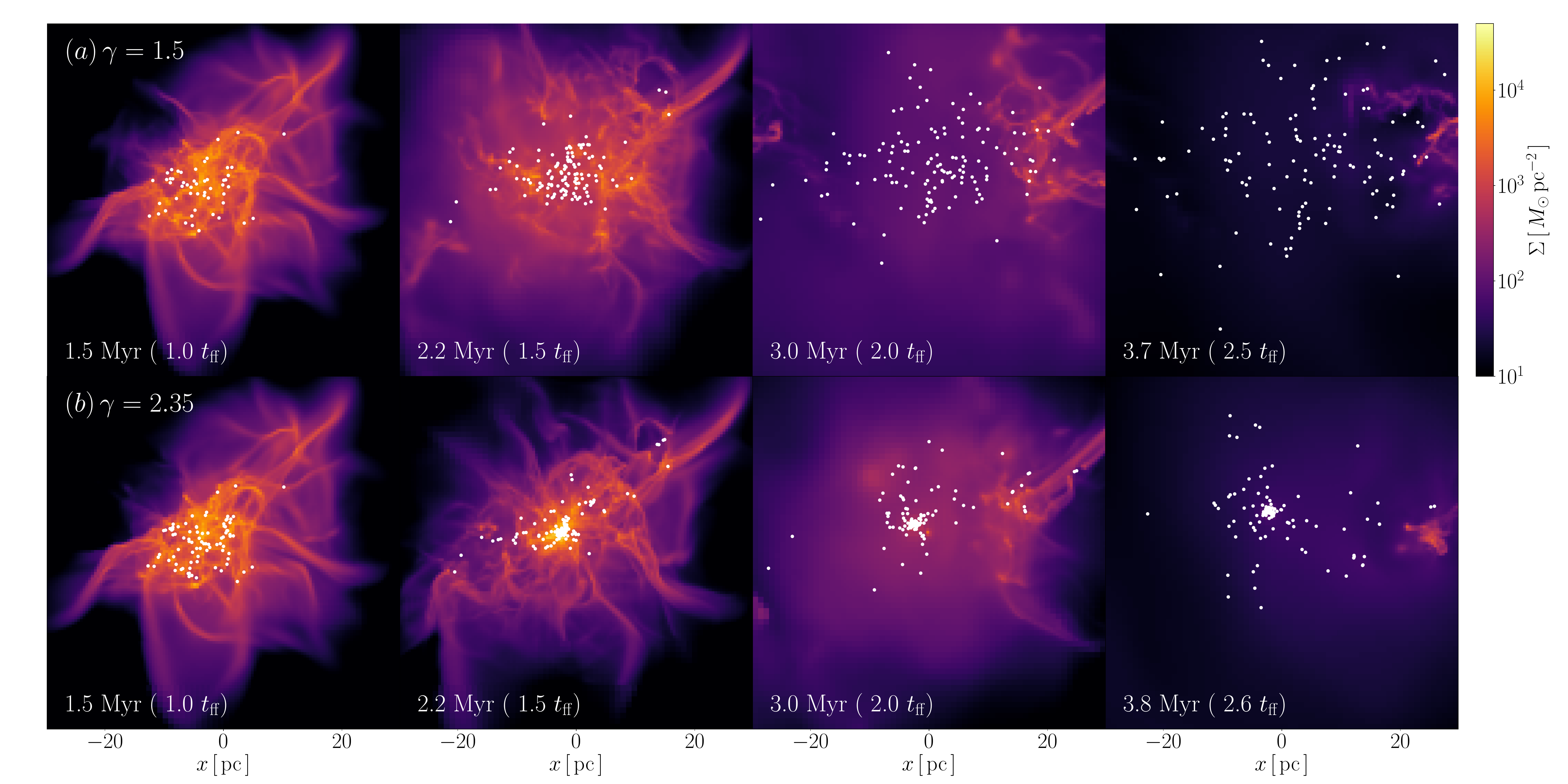}
    \end{center}
    \caption{
    Star cluster formation and cloud dispersion in the clouds with $M_{\rm cl} = 10^6~M_{\odot}$ and $\Sigma_{\rm cl} =800~M_{\odot}{\rm pc^{-2}}$. 
    The panels show the evolution of the surface densities.
    The top and bottom rows show the models with $\gamma = 1.5$ and $2.35$.
    The white dots represent the positions of stellar particles.
}   
    \label{fig:SNAP_SHOT2}
\end{figure*}

We present the star cluster formation in the case of the models with $(M_{\rm cl}, \Sigma_{\rm cl}) = (10^{6}~M_{\odot}, 800~M_{\odot})$.
Figure \ref{fig:SNAP_SHOT2} shows the time evolution of the gas surface densities with the top-heavy 
 IMF ($\gamma = 1.5$) and the standard IMF ($\gamma = 2.35$).
The entire evolution is almost the same until the elapsed time $t \sim 1.0~t_{\rm ff}$ where $t_{\rm ff}$ is the free-fall time of the cloud.
Once a few percent of gas is converted into stars ($t \sim 1.5~t_{\rm ff}$), gas clumps around stars start to be evacuated only in the case of the top-heavy IMF.
As a result, the gravitational potential is too shallow to bind the star cluster compactly, allowing the expansion of stellar distribution.
In this case, the star cluster has a stellar mass density lower than that of GCs.
On the other hand, in the case of the standard IMF, gas inflow and star formation continue against the stellar feedback even at $t \gtrsim 1.5~t_{\rm ff}$. Finally, the compact star cluster form at the center of the cloud, and its density is comparable to the observed GCs.

\begin{figure}
    \begin{center}
    	\includegraphics[width=\columnwidth]{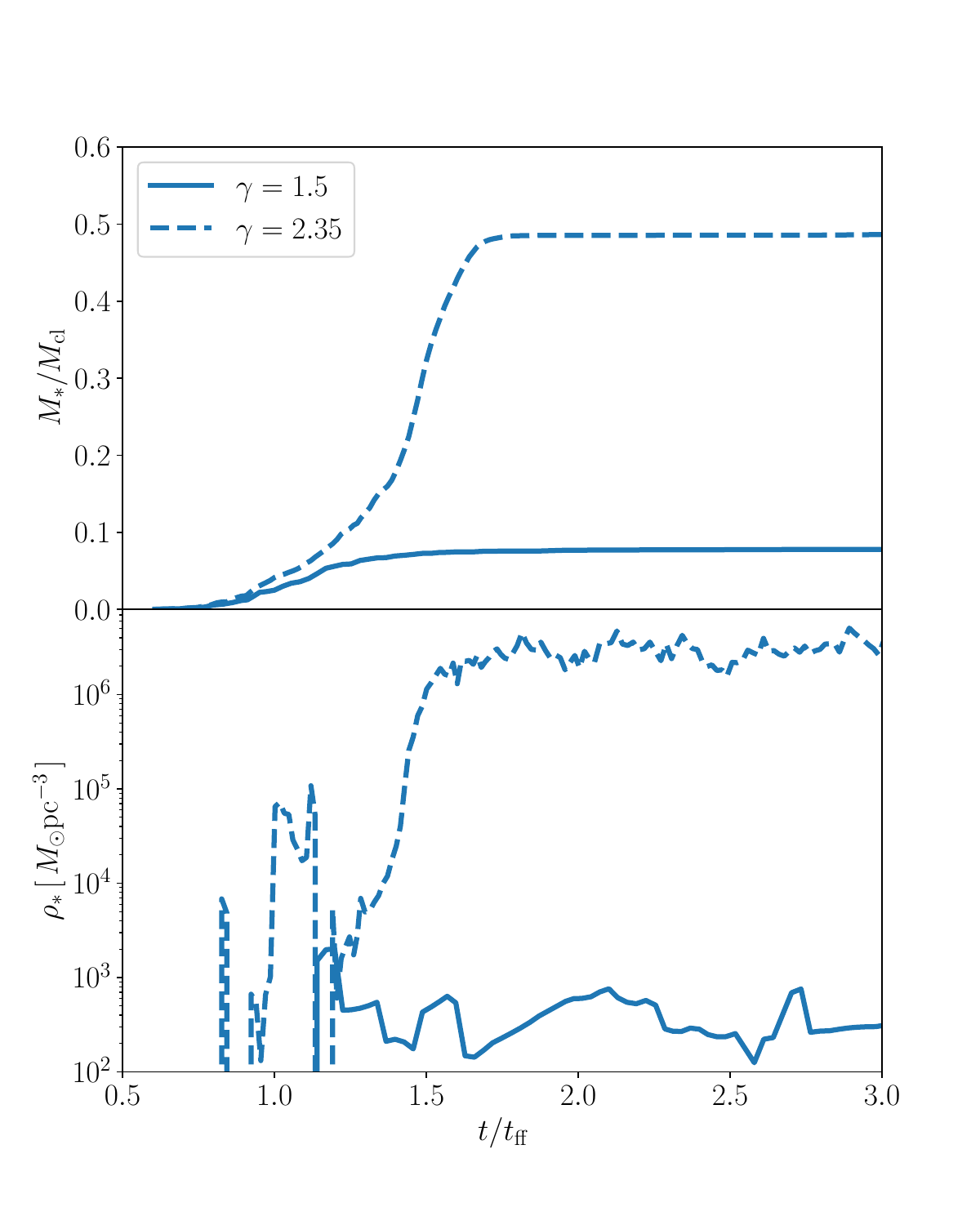}
    \end{center}
    \caption{
    Upper panel: the time evolution of the stellar mass in the cases with $(M_{\rm cl}, \Sigma_{\rm cl}) = (10^6~M_{\odot}, 800~M_{\odot}{\rm pc^{-2}})$.
    Lower panel: the time evolution of the stellar density.
    Each line represents the different IMF with $\gamma = 1.5$ (solid) and $2.35$ (dashed line).
}   
    \label{fig:mass_rho_star}
\end{figure}

Figure \ref{fig:mass_rho_star} shows the time evolution of the stellar mass and density in each model.
Here, we calculate the total mass and half-mass radius of the bound stars to obtain the stellar densities.
The details of the analytical method are described in \citetalias{2021MNRAS.506.5512F}.
The photoionization feedback suppresses the star formation significantly in the case of the top-heavy IMF ($\gamma = 1.5$).
The star formation efficiency (SFE) results in less than 10 percent.
In such a case, the stars are not bound by their own gravitational potential.
Thus, the resultant stellar density is lower than $10^3~M_{\odot} {\rm pc^{-3}}$.

In the case of the standard IMF ($\gamma = 2.35$), the star formation rate steeply increases at $t \gtrsim 1.3~t_{\rm ff}$.
In this phase, the high-density and gravitationally bound stellar core forms.
Finally, more than 40 percent of the gas is converted into stars, resulting in a massive compact star cluster. The stellar density shows $\sim 10^6~M_{\odot}{\rm pc^{-3}}$ which is similar to GCs.
We note that most gravitationally bound stars belong to the stellar cores. 
The density of the stellar core is almost the same as that of a star cluster, as shown in Table \ref
{Tab:results}.

\subsection{Dependence on IMF}\label{Sec:dependence_on_cloud_compact}

The large differences in the SFE and the stellar density are related to the stellar core formation as shown in Figure \ref{fig:SNAP_SHOT2}.
If the gas accretion continues against the photoionization feedback, the high-density stellar core forms at the center of the cloud.  
In \citetalias{2021MNRAS.506.5512F}, we analytically derive the threshold surface density for the stellar core formation as
\begin{align}
    \Sigma_{\rm cl} &> \Sigma_{\rm thr} = 670~M_{\odot}{\rm pc^{-2}} \left( \frac{M_{\rm cl}}{10^6~M_{\odot}} \right)^{-1/5} \left( \frac{s_*}{10^{47} ~M_{\odot}^{-1} s^{-1}} \right)^{2/5},
    \label{eq:thrsigma}
\end{align}
where $s_*$ is the number of ionizing photons per unit stellar mass.
Here, we consider the expanding shell model as in \citet{2002ApJ...566..302M} and \citet{2009ApJ...703.1352K}.
The deviation of $\Sigma_{\rm thr}$ is shown in Appendix \ref{section:shell_expansion_model}.
The cloud of $(M_{\rm cl}, \Sigma_{\rm cl}) = (10^6~M_{\odot}, 800~M_{\odot}{\rm pc^{-2}})$ satisfies this condition with the standard IMF. 
However, the threshold surface density increases up to $\Sigma_{\rm cl} > 1600~M_{\odot} {\rm pc^{-2}}$ in the case of the top-heavy IMF $(\gamma = 1.5)$.
Therefore, the massive and high-dense star clusters are difficult to form as the IMF is top-heavy.

\begin{figure}
    \begin{center}
    	\includegraphics[width=\columnwidth]{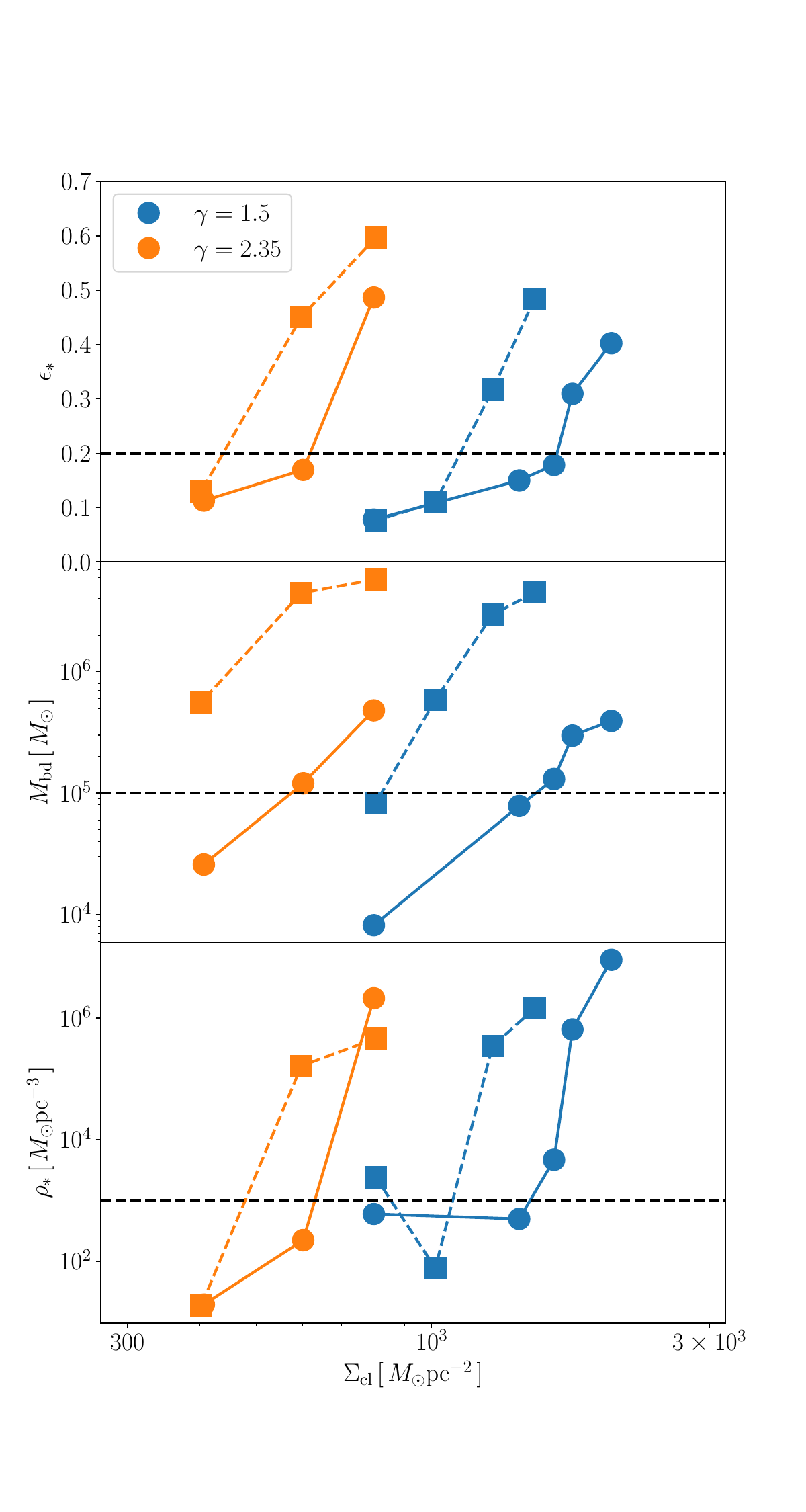}
    \end{center}
    \caption{
    Upper panel:  SFEs of clouds as a function of surface densities. The back dashed line represents $\epsilon_* = 0.2$.
    Lower panel: Stellar densities of star clusters.
    Each symbol represents the different cloud masses: $M_{\rm cl} = 10^6~M_{\odot}$ (circle) and $10^7~M_{\odot}$ (square). Each color shows the slope of the IMF: $\gamma = 1.5$ (blue) and $\gamma = 2.35$ (orange).
}   
    \label{fig:eps_rhostar}
\end{figure}

Figure \ref{fig:eps_rhostar} shows the dependencies of SFEs and the stellar densities on the surface density in the cases with the various cloud masses ($M_{\rm cl}=10^6~M_{\odot}$ and $10^7~M_{\odot}$) and the slope of the IMF ($\gamma = 1.5$ and $2.35$) as Tabel \ref{Tab:models}.
We estimate these values when the elapsed time of the simulations is $t=2~t_{\rm ff}$.
The SFE and the stellar densities increase with the higher surface densities.
The stellar core formation occurs when the SFE typically exceeds 0.15-0.2.
In such a case, the bound mass and the stellar densities are higher than $10^5~M_{\odot}$ and $>10^5~M_{\odot}{\rm pc^{-3}}$ which are similar to the properties of observed GCs.
With the top-heavy IMF ($\gamma = 1.5$), the core formation occurs at $\Sigma_{\rm cl} > 1600~M_{\odot}{\rm pc^{-2}}$ ($1000~M_{\odot}{\rm pc^{-2}}$) in the clouds with $M_{\rm cl} = 10^6~M_{\odot}$ ($10^7~M_{\odot}$).
In the case with the standard IMF ($\gamma = 2.35$), the threshold surface density decreases to $\Sigma_{\rm thr} = 750~M_{\odot}{\rm pc^{-2}}$ ($470~M_{\odot}{\rm pc^{-2}}$) with $M_{\rm cl} = 10^6~M_{\odot}$ ($10^7~M_{\odot}$) due to the decrease of the emissivity of the ionizing photons.
These results are consistent with the threshold surface density given as Equation \eqref{eq:thrsigma}.

\begin{figure*}
    \begin{center}
     \includegraphics[width=13cm]
     {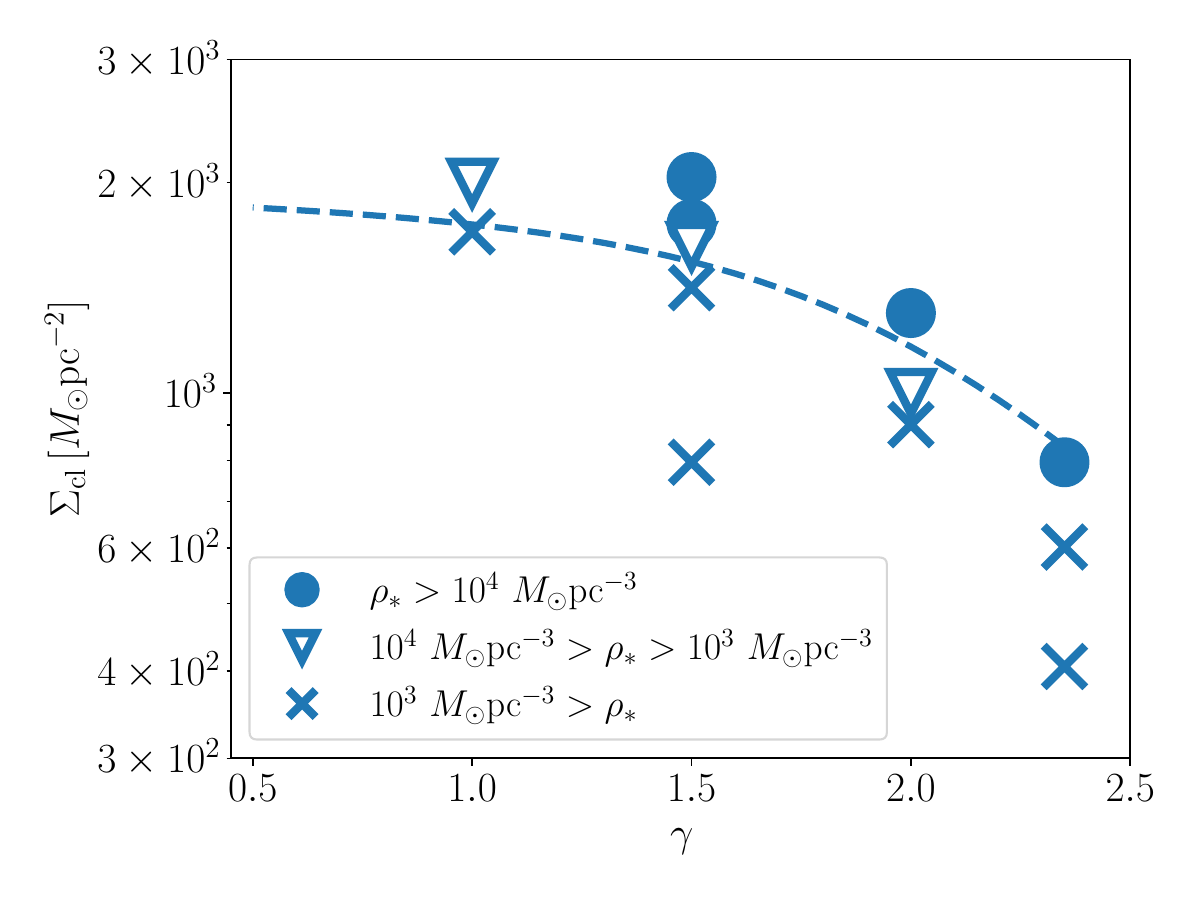}
    \end{center}
    \caption{The threshold surface density $(\Sigma_{\rm thr})$ for GC formation with $M_{\rm cl} = 10^{6}~M_{\odot}$ as a function of the slope of the IMF $(\gamma)$. 
    The dashed line shows the analytical result given as Eq. \eqref{eq:thrsigma}.
    The symbols indicate the stellar densities of star clusters.
    The circles, triangles, and crosses correspond to the cases with $\rho_* > 10^4~M_{\odot}{\rm pc^{-3}}$, $10^4~M_{\odot}{\rm pc^{-3}} > \rho_* > 10^3~M_{\odot}{\rm pc^{-3}}$, and $10^3~M_{\odot}{\rm pc^{-3}} > \rho_*$, respectively. 
    In the cases of triangles and circles, the bound stellar mass exceeds $10^5~M_{\odot}$.
    } 
    \label{fig:gam_sigma}
\end{figure*}

To investigate the dependency of the IMF slope on $\Sigma_{\rm thr}$, we additionally perform the models with $\gamma = 1$ and $2$ as Table \ref{Tab:models}.
Figure \ref{fig:gam_sigma} shows the numerical results of the stellar densities in the plane of $\gamma$ and $\Sigma_{\rm cl}$ with the cloud mass of $M_{\rm cl}=10^6~M_{\odot}$.
The triangles and circles represent the cases where the star cluster satisfies the properties of GC progenitors \citep[$M_{\rm bh} > 10^5~M_{\odot}$ and $\rho_* > 10^3~M_{\odot}{\rm pc^{-3}}$, ][]{2010ARA&A..48..431P}. 
Some GCs in the Milky Way have lower stellar density than this threshold value. However, star clusters gradually expand with time after their birth \citep[e.g.,][]{2009A&A...498L..37P, 2016ApJ...817....4F}.
Therefore, the stellar density $10^3~M_{\odot}{\rm pc^{-3}}$ can be reasonable condition for the GC formtaion.
The threshold surface density $\Sigma_{\rm thr}$ derived in the analytical model (eq. \ref{eq:thrsigma}) reproduces the simulation results well.
The threshold value is $750~M_{\odot}{\rm pc^{-2}}$ for $\gamma = 2.35$, and it increases as $\gamma$ decreases. At $\gamma \lesssim 1.5$, the threshold surface density is not sensitive to $\gamma$ because $s_{*}$ does not change with $\gamma$ significantly. Thus, we suggest that once a high-density cloud with $\Sigma \gtrsim 2\times 10^{3} M_{\odot}{\rm pc^{-2}}$ forms, GCs are likely to form even if the IMF is top-heavy ($\gamma \l 1.5$).

\section{Discussion}\label{Sec:discussion}

In this paper, we study the impacts of the IMF on the condition for GC formation in low-metallicity clouds.
The observations of the GCs in the Milky Way showed that there is the metallicity floor at $Z\sim 10^{-2.5}~Z_{\odot}$ \citep[e.g.,][]{2005A&A...439..997P, 2019MNRAS.487.1986B}.
This metallicity floor can be related to the formation sites and mechanisms of GCs.
\citet{2019MNRAS.486L..20K} indicated that the metallicity floor results from the mass-metallicity relation of their host galaxies. 
They showed that the galaxies with metallicities less than the floor are too small to form GCs.
\citet{2018MNRAS.475L.130A} pointed out that radiation pressure caused by the Ly $\alpha$ photos suppresses the bound star cluster in the low-metallicity environments ($Z\lesssim 10^{-2.5}~Z_{\odot}$).
We suggested that more compact clouds ($\Sigma_{\rm cl} > 10^3~M_{\odot}{\rm pc^{-2}}$) are nessesary for the GC formation at $Z\lesssim 10^{-3}~Z_{\odot}$ if the IMF is a top-heavy \citep{2021MNRAS.508.4175C}.
In such high surface density environments, the formed bound star cluster is more likely to be destroyed due to the interactions with another star-forming cloud \citep[e.g.,][]{1987degc.book.....S, 1995ApJ...438..702K, 2006MNRAS.371..793G, 1999ApJ...522..935G, 2012MNRAS.426.3008K}.
Besides, the star clusters with the top-heavy IMF cannot survive for a long time \citep[e.g.,][]{2020MNRAS.491..440W, 2021MNRAS.504.5778W}.
In such a star cluster, more remnant black holes (BHs) form and migrate to the center by kicking neighbor low-mass stars. 
These star clusters can be finally evaporated or become dark star clusters that only contain BHs \citep{2011ApJ...741L..12B}.
On the other hand, observations have discovered the GCs with metallicities lower than the metallicity floor in most GCs \citep[e.g.,][]{2020Sci...370..970L, 2020Natur.583..768W, 2022Natur.601...45M}.
Therefore, further studies are needed to connect the formation of low-metallicity GCs, their stellar IMF, and their survival rates in the early galaxies.

The formation sites of young massive star clusters have been observed frequently in the starburst and merger galaxies \citep[e.g.,][]{2018ApJ...869..126L, 2021PASJ...73..417T}.
The numerical simulations have indicated massive clouds could be induced with galaxy merger processes \citep[e.g.,][]{2018MNRAS.475.4252A, 2019ApJ...879L..18L}.
In these environments, the colliding flow with thermal or gravitational instability can be key to forming massive compact clouds \citep{2012ApJ...759...35I, 2020ApJ...905...95K, 2021ApJ...908....2M, 2020MNRAS.496L...1D, 2022MNRAS.509..954D}.
The above large-scale conditions can determine the initial condition of clouds as used in our work. We will connect the global structure in a galaxy and the star cluster formation in a cloud in future work.

In this work, we do not include the effects of magnetic fields.
Recent numerical studies have shown that magnetic fields have crucial roles, even in the first star formation \citep[e.g.,][]{2020MNRAS.497..336S, 2021MNRAS.503.2014S, 2022MNRAS.511.5042S, 2021MNRAS.505.4197S, 2022MNRAS.511.5042S, 2023MNRAS.519.3076S}.
Strong magnetic fields suppress star formation and decrease star formation efficiencies and rates in star cluster formation  \citep[e.g.,][]{2021ApJ...911..128K}.
The expansion velocity of H{\sc ii} regions can be lower with lower star formation rates.
Thus, GCs can be born in a cloud with lower surface density (see also Appendix \ref{section:shell_expansion_model}).
However, note that, if the duration of star formation is longer than the lifetime of massive stars, supernova feedback can suppress the GC formation. 
For further studies, we will include magnetic fields in future works.

We investigate the cases with $Z=10^{-3}~Z_{\odot}$ in this paper. 
At $Z \lesssim 10^{-2}~Z_{\odot}$, the temperature of H{\sc ii} regions does not change significantly \citep[e.g.,][]{2011piim.book.....D}, and
dust attenuation of UV photons is inefficient even for high-density filamentary structures  \citep{2020MNRAS.497.3830F}. 
Therefore, the GC formation is unlikely to be sensitive to the metallicity at $Z \lesssim 10^{-2}~Z_{\odot}$.
On the other hand, the temperature of ionizing gas decreases, and filaments are shielded by dust grains at higher metallicities. 
In such a case, the photoionization feedback is slightly more ineffective in suppressing star formation.
In addition, \citet{2022MNRAS.517.1313M} showed that infrared radiation pressure does not sufficiently decrease star formation efficiency even at solar metallicity.
Therefore, GCs can form more frequently in clouds with higher metallicity.

\section{Summary}\label{Sec:summary}

We have performed three-dimensional radiative-hydrodynamics  simulations of the formation of the low-metallicity globular clusters (GCs) with $Z=10^{-3}~Z_{\odot}$.
We have investigated the impacts of different initial mass functions (IMFs) on the properties of emergent star clusters.
Our simulations cover the various cloud masses and surface densities, $M_{\rm cl} = 10^6-10^7~M_{\odot}$ and $\Sigma_{\rm cl} =400-2000~M_{\odot}{\rm pc^{-2}}$.
Our findings are summarized as follows:

\begin{description}
\item[(i)] GCs form only in compact clouds with $\Sigma_{\rm cl} \sim 800~M_{\odot}{\rm pc^{-2}}$ for the cloud mass $M_{\rm cl}=10^6~M_{\odot}$ in the case of standard Salpeter-like IMF ($\gamma=2.35$).
With the top-heavy IMF of $\gamma = 1.5$, the threshold surface density ($\Sigma_{\rm thr}$) increases up to $\Sigma_{\rm thr} \sim 1600~M_{\odot}{\rm pc^{-2}}$ due to the higher emissivity of ionizing photons per unit stellar mass.
For $\gamma < 1.5$, $\Sigma_{\rm thr}$ saturates with $\Sigma_{\rm thr} \sim 2000~M_{\odot}{\rm pc^{-2}}$ because of saturation of the photon emissivity.

\item[(ii)] The simulation results about the threshold surface densities are reproduced well by a semi-analytical model updated from that derived in \citet{2021MNRAS.506.5512F}.
The threshold surface density mainly depends on the cloud mass and the emissivity of ionizing photons ($s_*$) as $\Sigma_{\rm cl} \propto M_{\rm cl}^{-1/5} s_{\rm *}^{2/5}$.

\end{description}

Theoretical studies predicted that GCs formed in dwarf galaxies in the early Universe
\citep[e.g.,][]{2016ApJ...831..204R, 2021MNRAS.500.4578M} due to the high Jeans mass \citep{1968ApJ...154..891P}, or thermal instability in low-metallicity gas \citep{1985ApJ...298...18F, 2018MNRAS.475.4252A}.
The collapse of the clouds under UV background radiation also induces the GC formation \citep{2009MNRAS.397.1338H, 2016MNRAS.463.2849A}.
Thanks to \textsc{James Webb Space Telescope} (JWST), star-formation sites in high-redshift galaxies have been observed directly \citep[e.g.,][]{2022arXiv220800520V}. Also, recent observations suggested the physical properties of the gas, e.g., density, temperature, and metallicity in star-forming regions of high-redshift galaxies \citep[e.g.,][]{2023arXiv230106811I}. 
Combining the GC formation condition obtained in this study with the observations, we will study the origin of GCs in future work.
Recently, \citet{2022ApJ...938L..10I} indicated that star-formation efficiencies (SFEs) can be higher than $\sim 0.1-0.3$ in galaxies at $z > 10$ from observed UV luminosity functions
\citep[e.g.,][]{2022ApJ...929....1H, 2022ApJS..259...20H, 2022arXiv220712474F}.
These high SFEs satisfy a condition for the GC formation.
With upcoming observations, we will study the GC formation at $z \gtrsim 10$ with cosmological simulations of galaxy formation \citep[e.g.,][]{2021MNRAS.508.3226A, 2022MNRAS.509.4037Y}.

\section*{Acknowledgements}
The authors wish to express their cordial thanks to Profs. Masayuki Umemura and Ken Ohsuga for their continual interest, advice, and encouragement.
We appreciate Tomoaki Matsumoto for his great contribution to code development.
We would like to thank Takashi Hosokawa for useful discussions and comments.
The numerical simulations were performed on the Cray XC50 (Aterui II) at the Center for Computational Astrophysics of National Astronomical Observatory of Japan and Yukawa-21 at Yukawa Institute for Theoretical Physics at Kyoto University.
This work is supported in part by MEXT/JSPS KAKENHI Grant Number 17H04827, 20H04724, 21H04489 (HY), NAOJ ALMA Scientific Research Grant Numbers 2019-11A, JST FOREST Program, Grant Number JP-MJFR202Z, and Astro Biology Center Project research AB041008 (HY). 

\section*{Data Availability}

The data underlying this article will be shared on reasonable request to the corresponding author.



\bibliographystyle{mnras}
\input{main.bbl}




\appendix

\section{Threshold surface density of globular cluster formation}\label{section:shell_expansion_model}

We summarize the underlying physics in massive star cluster formation under radiative feedback as described in \citet{2021MNRAS.506.5512F}.
We consider an expanding shell around an H{\sc ii} region.
The equation of motion is described as \citep{2002ApJ...566..302M, 2009ApJ...703.1352K}
\begin{align}
    \frac{d}{dt} \left( M_{\rm sh} \dot r_{\rm sh} \right) = 4 \pi r_{\rm sh}^2 \rho_{\rm i} c_{\rm i}^2, \label{eq:shell_eom}
\end{align} 
where $\rho_{\rm i}$ and $c_{\rm i}$ are the density and the sound speed in H{\sc ii} regions.
Here, we consider the contribution from the thermal pressure of ionized gas alone and ignore other effects, such as radiation pressure and gravitational force from stars.
When the photon production rate balances with the total recombination rate in the H{\sc ii} region, the number density of H{\sc ii} regions is evaluated as 
\begin{align}
    n_{\rm i} = \left( \frac{\rho_i c_{i}^2}{k_{\rm B} T_{\rm i}} \right) = \left( \frac{3 S_{\rm ion}}{4 \pi r_{\rm sh}^3 \alpha_{\rm B}} \right)^{1/2}, \label{eq:number_density_inHII}
\end{align}
where $T_{\rm i}$, $S_{\rm ion}$, and $\alpha_{\rm B}$ are the temperature of ionized gas, the emissivity of ionizing photons, and the recombination coefficient $\alpha_{\rm B} = 2.6 \times 10^{-13} (T_{\rm i}/10^4~{\rm K})^{-0.8} \, {\rm cm^{3}s^{-1}}$ \citep{1989agna.book.....O}.
Here, we assume that the gas is converted into stars with the constant conversion rates $\epsilon_*$ after the shell passes.
The mass of expanding shell and the formed stars inside the shell are given as 
\begin{align}
    M_{\rm sh} = M_{\rm cl} (1-\epsilon_*) (r_{\rm sh}/R_{\rm cl})^3, \label{eq:shellmass}
\end{align}
and 
\begin{align}
    M_* = M_{\rm cl} \epsilon_* (r_{\rm sh}/R_{\rm cl})^3. \label{eq:starmass}
\end{align}
The emissivity of ionizing photons is estimated as
\begin{align}
    S_{\rm ion} = s_* M_{*}, \label{eq:sion}
\end{align}
where $s_*$ is the emissivity per unit mass.
Here, we adopt the following dimensionless parameters 
\begin{align}
    x = r_{\rm sh}/R_{\rm cl}, \label{eq:nondimensional1}
\end{align}
and
\begin{align}
    \tau =  \sqrt{3} t/t_{\rm HII}, \label{eq:nondimensional2}
\end{align}
where 
\begin{align}
    t_{\rm HII} = \left( \frac{1-\epsilon_*}{\epsilon_*} \right)^{1/4} \left( \frac{3 \alpha_{\rm B}}{4 s_* k_{\rm B}^2 T_{\rm i}^2} \right)^{1/4} \left( \frac{M_{\rm cl}^3}{\pi^3 \Sigma_{\rm cl}} \right)^{1/8}, \label{eq:tHII}
\end{align}
Substituting equation \eqref{eq:number_density_inHII}, \eqref{eq:nondimensional1}, \eqref{eq:nondimensional2} into equation \eqref{eq:shell_eom}, the equation of shell motion is rewritten as 
\begin{align}
    \frac{d}{d \tau} \left(x^3 \dot x \right) = x^2. \label{eq:eqmotion_dimensionless}
\end{align}
The self-similar solution of equation \eqref{eq:eqmotion_dimensionless} is found as $x = \tau / \sqrt{3}$ and $\dot x = 1/\sqrt{3}$ \citep{2021MNRAS.506.5512F}.
Thus, the shell arrives at the outer edge of the cloud within the timescale of $t_{\rm HII}$.
We assume that the shell expansion time of $t_{\rm HII}$ is equal to the duration time of the star formation.
The total stellar mass formed inside the cloud is given as 
\begin{align}
    M_* = \epsilon_* M_{\rm cl} = \dot M_* t_{\rm HII}. \label{eq:total_stellar_mass}
\end{align}
The star formation rates are expressed as 
\begin{align}
    \dot M_* = \epsilon_{\rm ff} \frac{M_{\rm cl}}{t_{\rm ff}}, \label{eq:star_formation_rates}
\end{align}
where $t_{\rm ff} = \sqrt{3 \pi / (32 G \rho)}$ is the free-fall time of the cloud.
Substituting equations \eqref{eq:tHII} and \eqref{eq:star_formation_rates} into equation \eqref{eq:total_stellar_mass}, we obtain the SFE as
\begin{align}
    \epsilon_* \simeq 0.07 & \left( \frac{\epsilon_{\rm ff}}{0.03} \right)^{4/5}  \left( \frac{\Sigma_{\rm cl}}{10^3~M_{\odot}{\rm pc^{-2}}} \right)^{1/2} \left( \frac{M_{\rm cl}}{10^6~M_{\odot}} \right)^{1/10} \nonumber \\
    & \left( \frac{T_{\rm i}}{3 \times 10^4 ~{\rm K}} \right)^{-14/25} \left( \frac{s_*}{5.8 \times 10^{46} ~M_{\odot}^{-1} s^{-1}} \right)^{-1/5} \label{eq:sfe}
\end{align}
The shell expansion velocity does not depend on the radial position.
With the SFE given as equation \eqref{eq:sfe}, the shell velocity is calculated as
\begin{align}
    v_{\rm exp} &= \frac{R_{\rm cl}}{t_{\rm HII}} \nonumber \\ 
    &\simeq 5.8~{\rm km/s} \left( \frac{\epsilon_{\rm ff}}{0.03} \right)^{1/5} \left( \frac{\Sigma_{\rm cl}}{10^3~M_{\odot}{\rm pc^{-2}}} \right)^{-1/4}  \left( \frac{M_{\rm cl}}{10^6~M_{\odot}} \right)^{3/20} \nonumber \\
    & \hspace{2cm} \left( \frac{T_{\rm i}}{3 \times 10^4~{\rm K}} \right)^{14/25} \left( \frac{s_*}{5.8 \times 10^{46} ~M_{\odot}^{-1} s^{-1}} \right)^{1/5}. \label{eq:velocity_of_HIIregions}
\end{align}

The formed stars start to be bound when the total stellar mass exceeds 0.1 times the cloud mass \citep{2021MNRAS.506.5512F}.
At this epoch, ionizing fronts propagate beyond the star cluster.
When the thermal pressure on the shell overcomes the gravitational force, the gas around the star clusters is dispersed.
On the other hand, the dense gas remains, and H{\sc ii} regions are confined if the gravitational force is strong enough to hold the ionized gas.
In such a case, the gas keeps accreting onto the star cluster and contributes to further star formation. 
As a result, the high-density stellar core forms.
Assuming that the expanding shell around the star cluster has the same velocity as the semi-analytical solution of equation \eqref{eq:velocity_of_HIIregions}, the condition for binding the shell is given as
\begin{align}
   v_{\rm exp} < v_{\rm esc} = \sqrt{2GM_{\rm core}}{R_{\rm core}}, \label{eq:formation_stellar_core}
\end{align}
where $M_{\rm core}$ and $R_{\rm core}$ are the core mass and radius.
These values are typically $R_{\rm core} \sim 0.1 R_{\rm cl}$ and $M_{\rm core} \sim 10^{-2}~M_{\rm cl}$.
We obtain the conditions for the formation of a massive star cluster as 
\begin{align}
    \Sigma_{\rm cl} > \Sigma_{\rm thr} &= 750~M_{\odot}{\rm pc^{-2}} \left( \frac{\epsilon_{\rm ff}}{0.03} \right)^{2/5}  \left( \frac{M_{\rm cl}}{10^6~M_{\odot}} \right)^{-1/5} \nonumber \\ 
    & \hspace{5mm} \left( \frac{T_{\rm i}}{2.5 \times 10^4~{\rm K}} \right)^{28/25} \left( \frac{s_*}{1.1 \times 10^{47} ~M_{\odot}^{-1} s^{-1}} \right)^{2/5}. \label{eq:sig_thr}
\end{align}
In this study, we do not include magnetic fields. 
If strong magnetic fields exist, star formation rates decrease \citep[e.g.,][]{2021ApJ...911..128K}.
The threshold value of equation \eqref{eq:sig_thr} depends on star formation rates as $\Sigma_{\rm thr} \propto \epsilon_{\rm ff}^{2/5}$.
If $ \epsilon_{\rm ff}$ is reduced to half by magnetic fields, the condition of young massive star cluster formation is relaxed as $\Sigma_{\rm thr}\sim 570~M_{\odot}{\rm pc^{-2}}$.

\begin{figure}
    \begin{center}
    	\includegraphics[width=\columnwidth]{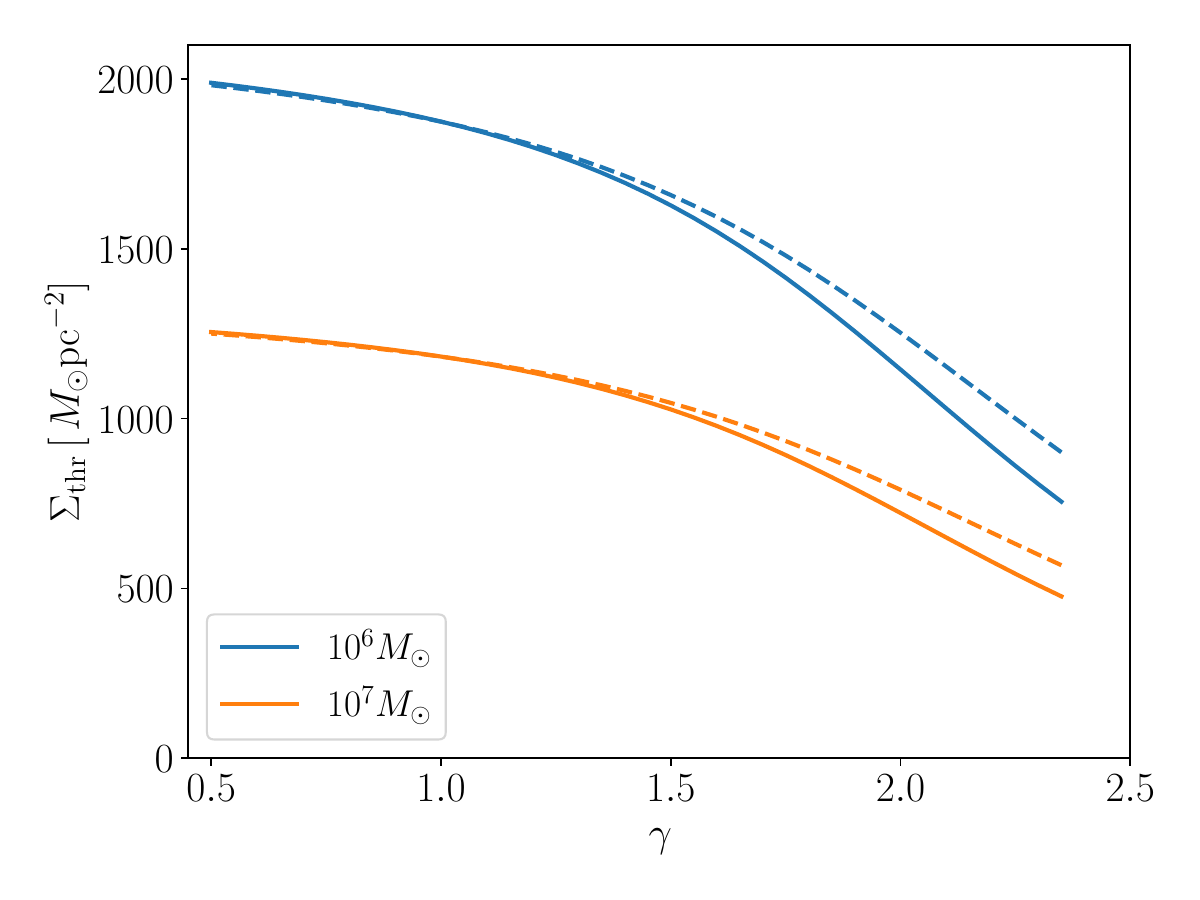}
    \end{center}
    \caption{
    The threshold surface density ($\Sigma_{\rm thr}$) as the function of the slope of the IMF ($\gamma$). Each line shows the cases with the cloud of $M_{\rm cl}=10^6~M_{\odot}$ (blue) and $10^7~M_{\odot}$ (orange).
    The solid (dashed) line shows the cases that the maximum stellar mass is $300~M_{\odot}$ ($150~M_{\odot}$).}   
    \label{fig:gam_sstar}
\end{figure}

Combining the emissivity of ionizing photons as shown in Figure \ref{fig:sstar_ion_gam}, we calculate the threshold surface densities as the function of the power-law index of the IMF ($\gamma$).
Figure \ref{fig:gam_sstar} shows the values of $\Sigma_{\rm thr}$ as a function of $\gamma$ in the clouds with $M_{\rm cl} = 10^6~M_{\odot}$ and $10^7~M_{\odot}$.
In both cases, the threshold surface densities are less than $10^3~M_{\odot}{\rm pc^{-2}}$ for the standard IMF ($\gamma \sim 2.35$).
At $\gamma \gtrsim 1.5$, the values of $\Sigma_{\rm thr}$ are larger than $\sim 10^{3}~M_{\odot}{\rm pc^{-2}}$ of which clouds are rare in  observations of local galaxies \citep[e.g.,][]{2010ApJ...723..492R}.
We also consider the dependence of $\Sigma_{\rm thr}$ on the maximum stellar mass ($M_{\rm max}$) of the IMF.
The dot and dashed lines in Figure \ref{fig:gam_sstar} show the cases that the maximum stellar masses are $M_{\rm max} = 300~M_{\odot}$ and $150~M_{\odot}$ with the Salpeter IMF.
At $\gamma < 1.5$, the values of $\Sigma_{\rm thr}$ depend on the maximum stellar mass.
However, the threshold surface density does not vary with the maximum stellar mass at $\gamma > 1.5$.


\bsp	
\label{lastpage}
\end{document}